\documentclass{PoS}
\usepackage{amsmath}
\title{Phenomenology of Irreversible Processes from Gravity}

\ShortTitle{Phenomenology of Irreversible Processes from Gravity}

\author{\speaker{Ayan Mukhopadhyay}\thanks{We thank B. Pioline and R. Gopakumar for comments on the manuscipt.}\\
        Laboratoire de Physique Th\'eorique et Hautes
Energies (LPTHE)\\
Universit\'e Pierre et Marie Curie -- Paris 6; CNRS UMR
7589 \\ 
Tour 13-14, 4$^{\grave{e}me}$ \'etage, Boite 126, 4 Place
Jussieu, 75252 Paris Cedex 05, France\\
        E-mail: \email{ayan@lpthe.jussieu.fr}}

\author{Ramakrishnan Iyer\\
     Department of Physics and Astronomy, University
of Southern California\\
Los Angeles, California 90089-0484, USA \\
       E-mail: \email{ramaiyer@usc.edu}}

\abstract{We propose that the space-time evolution of strongly coupled matter formed by ultra-relativistic heavy ion collisions can be modelled by phenomenological equations involving the energy-momentum tensor and conserved currents alone. These equations can describe the late stage of local thermal and chemical equilibration of the matter formed after collisions, and its subsequent transition to hydrodynamic expansion in an unified framework. The full set of equations include local energy, momentum and charge conservation; but also additional equations for evolution of non-equilibrium variables. These equations with precisely determined phenomenological parameters can be obtained by the AdS/CFT correspondence. On the gravity side of this correspondence, for vanishing chemical potentials, these phenomenological equations give all solutions of pure gravity in AdS which have regular future horizons. We also discuss field-theoretic grounds for validity of these phenomenological equations.}

\FullConference{The 2011 Europhysics Conference on High Energy Physics-HEP 2011,\\
		July 21-27, 2011\\
		Grenoble, Rhône-Alpes France}

\begin{document}

\section{Introduction and outlook}

Modelling the space-time evolution of the matter created by ultra-relativistic heavy ion collisions is a great challenge. Experiments at RHIC suggest the validity of the following picture \cite{Florkowski} : (i) a large fraction of the initial kinetic energy of the colliding ions is thermalized astonishingly fast (in time $\leq$ 1 fm) forming a locally equilibrated hot and dense fireball parametrized by a profile of the hydrodynamic variables - namely the temperature, four-velocity and chemical potential fields, (ii) the strongly interacting fireball undergoes hydrodynamic expansion \footnote{This is usually modelled by the relativistic Navier-Stokes equation.}, and (iii) the initial transverse hydrodynamic flow at the time of local thermal equilibration in most cases vanishes. Most of the data at RHIC is in good agreement with this simplistic picture especially in the mid-rapidity region, i.e. for the most central collisions at the highest beam energy of $\sqrt{s_{NN}} = 200$ GeV. However, despite the success in explaining the transverse momentum spectra of hadrons, the elliptic flow coefficient, etc., this does not reproduce pion interferometric data like HBT radii leading to the well-known RHIC HBT puzzle.

It is necessary to have a better phenomenological model for the space-time evolution of the fireball to explain the data completely, and also to reduce theoretical uncertainties. The best theoretical tool at hand for studying evolution of strongly coupled matter of gauge theories in real time is the AdS/CFT correspondence. It is well-known that the AdS/CFT correspondence gives $\eta/s = 1/4\pi$ \cite{Policastro}, while the current analysis of experimental data suggests $1 < 4\pi\Big(\eta/s\Big) < 2.5$ for temperatures probed at RHIC \cite{Heinz}. In fact, the AdS/CFT correspondence also predicts systematic hydrodynamic corrections to the Navier-Stokes equation (for a review see \cite{Janik}).

Here we will propose that the AdS/CFT correspondence can be used to develop a complete phenomenology for the evolution of the strongly coupled matter, describing both the late stages of local thermalization and the subsequent hydrodynamic expansion in an unified framework. These phenomenological equations involve a closed set of equations for evolution of the energy-momentum tensor and the baryon number  charge current alone. 

The advantage of our proposal is that there is a very natural way to connect the expansion of the fireball with any model which describes the early stages of the collision process, as for instance the parton cascade model \cite{Geiger}. All that is needed is to match the evolution of the energy-momentum tensor and conserved charge currents before and after the matter enters in the strongly coupled phase of evolution. The entry into the strongly coupled phase can be traced through the temperature field given by the energy-momentum tensor itself as we will discuss later. Importantly, the matching with the initial regime does not require the energy-momentum tensor to be hydrodynamic.

\section{Field-theoretic grounds}

Before we use the AdS/CFT correspondence to obtain the phenomenological equations for strongly coupled irreversible processes, it will be useful to see on what field-theoretic grounds we can use only the energy-momentum tensor and conserved charge currents to construct these phenomenological equations.  To keep the discussion simple, we will assume the baryon chemical potential is zero all throughout the evolution. It is in principle straightforward to include the charge currents.

It is known that the Boltzmann equation with a collision kernel determined by two body parton scattering and fragmentation processes can capture all perturbative non-equilibrium processes in non-Abelian gauge theories \cite{Arnold}. It can be shown that the relativistic semiclassical Boltzmann equation has special solutions, named "conservative solutions",  and are such that they can be determined by the energy-momentum tensor alone \cite{myself1}. In this approximation, the energy-momentum tensor is parameterized by the first ten velocity moments of the parton distribution functions in phase space. It can be shown that all the higher velocity moments of the parton distribution functions have special algebraic solutions of their equations of motion, such that they are algebraic functions of the energy-momentum tensor and its derivatives, having no independent dynamics. These algebraic functions can be expanded systematically in the hydrodynamic derivative expansion and the non-hydrodynamic amplitude expansion, to be discussed later.  The energy-momentum tensor follows a closed system of equations of motion, which includes the conservation of energy and momentum, but also equations for evolutions of the non-equilibrium variables giving its complete evolution. These equations can be obtained from the Boltzmann equation, and all phenomenological parameters including transport coefficients can be obtained from the collision kernel. Obviously, any solution of these equations can be lifted to a unique full solution of the Boltzmann equation.

Furthermore, any arbitrary solution of the Boltzmann equation at sufficiently late time can be approximated by an appropriate conservative solution, and  any conservative solution becomes purely hydrodynamic at late time \cite{myself1}. The purely hydrodynamic solutions of Boltzmann equation are known as "normal solutions" in literature \cite{Chapman}.  So, we indeed get a field-theoretic justification for using phenomenological equations involving the energy-momentum tensor alone in order to describe general irreversible processes. Also these phenomenological equations describe transition to hydrodynamic regime.

\section{General phenomenology and AdS/CFT}

In the strong coupling regime, the phenomenological equations of the evolution of the energy-momentum tensor should be obtained from gravity. Using consistent truncations of equations of motion of gravity, it can be shown that any solution of Einstein's equation with negative cosmological constant maps to a non-equilibrium state in the gauge theory via AdS/CFT correspondence, provided the solution has a regular future horizon. Furthermore, these solutions are determined uniquely by the boundary energy-momentum tensor \cite{myself2}, which by the AdS/CFT dictionary maps to the expectation value of the energy-momentum tensor in the dual state. It can be expected that when the energy-momentum tensor follows phenomenological equations with right values of the phenomenological parameters, the solutions in gravity will have regular future horizons - thus gravity should determine uniquely all phenomenological parameters.

To construct the general phenomenological equations for the energy-momentum tensor, we do not need either the Boltzmann equation or gravity, the latter are required only for determining the phenomenological parameters \cite{myself1, myself3}. Any arbitrary energy-momentum tensor can be written in the Landau-Lifshitz decomposition as :
\begin{eqnarray}\label{def}
t_{\mu\nu}(x) &=& e\left(T(x)\right) u_\mu (x) u_\nu (x) + p\left(T(x)\right)  P_{\mu\nu}(x) + \pi_{\mu\nu}(x), \ \text{with}\\\nonumber  \ P_{\mu\nu}(x) &=& u_\mu(x)u_\nu(x) +\eta_{\mu\nu} \ \text{and} \ u^\mu(x)\pi_{\mu\nu}(x) = 0.
\end{eqnarray}
The four-velocity $u^\mu(x)$ is the local velocity of energy-transport. The non-equilibrium part $\pi_{\mu\nu}(x)$ therefore is orthogonal to the four-velocity field and thus have six independent components;  so including the velocity and temperature fields we have ten independent variables. Conformality requires $e(x) = 3p(x)$ and $\pi_{\mu\nu}(x)$ to be traceless. We will also normalize the temperature such that $e(x) = (3/4)\cdot \left(\pi T(x)\right)^4$.  

The constraints of Einstein's equations automatically gives the conservation of energy and momentum :
\begin{equation}\label{cons}
\partial^\mu t_{\mu\nu} = 0.
\end{equation}

Without loss of generality, $\pi_{\mu\nu}$ can be split into a purely hydrodynamic part $\pi_{\mu\nu}^{(h)}$ and a non-hydrodynamic part $\pi_{\mu\nu}^{(nh)}$ which cannot be determined by hydrodynamic variables alone. Thus,
\begin{equation}\label{spl}
\pi_{\mu\nu} = \pi_{\mu\nu}^{(h)}+ \pi_{\mu\nu}^{(nh)}.
\end{equation}
The hydrodynamic part, $\pi_{\mu\nu}^{(h)}$ has a purely hydrodynamic derivative expansion, the expansion parameter $\epsilon$ is the ratio of the typical length scale of variation to the mean-free path. Requiring conformal invariance, and using the AdS/CFT correspondence to obtain the transport coefficients, we get up to second order in derivative expansion \cite{Janik},
\begin{eqnarray}\label{soh}
\pi_{\mu \nu}^{(h)} &=& - 2(\pi T)^3 \sigma_{\mu \nu} +
(2 - \ln 2)(\pi T)^2 \mathcal{D}\sigma_{\mu \nu} + 2 (\pi T)^2
\left(\sigma_{\mu}^{\phantom{\mu}\alpha}\sigma_{\alpha\nu} -
\frac{1}{3}P_{\mu \nu} \sigma_{\alpha \beta}\sigma^{\alpha
\beta}\right) \\\nonumber &&+\ln 2 (\pi T)^2
(\sigma_{\mu}^{\phantom{\mu}\alpha}\omega_{\alpha\nu}
+\sigma_{\nu}^{\phantom{\nu}\alpha}\omega_{\alpha\mu} ) +
O(\epsilon^3),
\end{eqnarray}
where $\sigma_{\mu\nu}$ is the shear-stress tensor, $\omega_{\mu\nu}$ is the velocity-vortex, and $\mathcal{D}$ is the Weyl-covariant convective derivative. 

The non-hydrodynamic part $\pi_{\mu\nu}^{(nh)}$ has an additional amplitude parameter $\delta$, which is the ratio of the typical non-hydrodynamic shear-stress to the equilibrium pressure. However, unlike the hydrodynamic variables, in a local inertial frame where the energy flow vanishes, close to equilibrium $\pi_{\mu\nu}^{(nh)}$ is slowly varyiing in space but not in time. So, at every order in amplitude expansion we must sum over all time-derivatives, or to state in a Lorentz and Weyl covariant manner - all Weyl-covariant convective derivatives $\mathcal{D}$. Furthermore, it should be possible to set consistently $\pi_{\mu\nu}^{(nh)}$ to zero as we know that the purely hydrodynamic sector exists in both the Boltzmann equation and gravity. Putting all these requirements together, and expanding both in $\epsilon$ and $\delta$
we get the most general Weyl and Lorentz covariant phenomenological equation for $\pi_{\mu\nu}^{(nh)}$ \cite{myself3}:
\begin{eqnarray}\label{rcg}
\left(\displaystyle\sum\limits_{n=0}^{\infty}D_{R}^{(1,n)}(\pi T)^n
\mathcal{D}^n\right)\pi_{\mu \nu}^{(nh)} &=&
\frac{(\pi T)\lambda_1 }{2}
\left(\pi_{\mu}^{(nh)\alpha}\sigma_{\alpha\nu}
+\pi_{\nu}^{(nh)\alpha}\sigma_{\alpha\mu}- \frac{2}{3}P_{\mu
\nu} \pi_{\alpha \beta}^{(nh)}\sigma^{\alpha \beta}\right)
\\\nonumber
&&+\frac{(\pi T)\lambda_2 }{2}
\left(\pi_{\mu}^{(nh)\alpha}\omega_{\alpha\nu}
+\pi_{\nu}^{(nh)\alpha}\omega_{\alpha\mu}\right) \\\nonumber
&&- (\pi T)^{4}\displaystyle\sum\limits_{n=0}^{\infty}\displaystyle\sum\limits_{\substack{m=0 \\ n+m \ is \ even}}^{n}D_R^{(2,n,m)}(\pi T)^{n}\displaystyle\sum\limits_{\substack{a,b=0 \\ a+b=n \\ |a-b|=m}}^{n}\Bigg[\mathcal{D}^a
\pi_{\mu}^{(nh)\alpha}\mathcal{D}^b
\pi_{\alpha\nu}^{(nh)}\\\nonumber &&\qquad\qquad\qquad\qquad\qquad\qquad\qquad  -\frac{1}{3}P_{\mu\nu} \ \mathcal{D}^{a}
\pi_{\alpha\beta}^{(nh)}\mathcal{D}^b \pi^{(nh)\alpha\beta}\Bigg]\\\nonumber
&&+ O(\epsilon^2\delta, \epsilon\delta^2 ,\delta^3).
\end{eqnarray}

It is very hard to give a general proof of validity of this equation in gravity, nevertheless it has been shown that it reproduces the gravity solutions dual to homogeneous relaxation \cite{myself3}. In such cases, the hydrodynamic variables are constant in space and time, while $\pi_{\mu\nu}^{(nh)}$ is spatially homogeneous. Such solutions also exist in the Boltzmann equation. It describes well the transition to local equilibrium. It has been found that the future horizon is regular, provided \emph{all convective derivatives are summed over at each order in the amplitude expansion} as expected. Furthermore, we obtain a complicated recursion relation for the phenomenological parameters $D_{R}^{(1,n)}$ and $D_R^{(2,n,m)}$, the first few terms being \cite{myself3}:
\begin{equation}
D_R^{(1,0)}= -1, \ \  D_R^{(1,1)}=-(\pi/2) - (1/4) \ \ln 2, \ \ \text{etc.;} \quad D_R^{(2,0,0)} = 1/2, \text{etc.}
\end{equation}

In order to obtain other parameters in (\ref{rcg}) like $\lambda_1$, $\lambda_2$, etc. it will be necessary to consider more general inhomogeneous configurations. Proving that the eqs. (\ref{def}), (\ref{cons}), (\ref{spl}), (\ref{soh}) and (\ref{rcg}) give all solutions of pure gravity in AdS with regular future horizons for  right phenomenological parameters, which have been partially determined here, will give us further confidence in these phenomenological equations \footnote{In practice, one needs to use these equations in a coordinate system better adapted for the late equilibrium state of the fireball which is an ideal fluid undergoing boost-invariant expansion. This coordinate system comprises of the proper time coordinate $\tau$ of the late time expansion, the coordinate $y$ parameterizing rapidity, and the two transverse coordinates $x^1$ and $x^2$. } .

\end{document}